\documentclass[journal]{IEEEtran}
\usepackage{cite}
\usepackage{amsmath,amssymb,amsfonts}
\usepackage{algorithmic}
\usepackage{graphicx}

\usepackage{diagbox}
\usepackage{multirow}
\usepackage{booktabs,array,xcolor,colortbl,multicol,epsfig,graphicx,subfigure}
\usepackage{amsmath,amsthm,amsfonts,amssymb,bm}
\usepackage[vlined,boxed,commentsnumbered,ruled,linesnumbered]{algorithm2e}
\usepackage{caption}
\usepackage{colortbl,accents}
\usepackage{subfigure}
\usepackage{caption}
\usepackage{color}

\usepackage{graphicx} 
\usepackage{epstopdf}

\usepackage{textcomp}

\begin{document}

\title{Modeling Event Propagation via Graph Biased Temporal Point Process}
\author{Weichang~Wu,
        Huanxi Liu$^*$,
        Xiaohu Zhang,
        Yu Liu,
        and~Hongyuan~Zha
\thanks{W. Wu, H. Liu and H. Zha are with MoE Key Lab of Artificial Intelligence, AI Institute, Shanghai Jiao Tong University, Shanghai, China. E-mail: \{blade091, lhxsjtu, zhasjtu\}@sjtu.edu.cn. *Corresponding author.}
\thanks{X. Zhang and Y. Liu are with China Telecom BestPay Co., Ltd., Shanghai, China. E-mail: \{zhangxiaohu, liuyu\}@bestpay.com.cn}}

\maketitle
\begin{abstract}
	Temporal point process is widely used for sequential data modeling. In this paper, we focus on the problem of modeling sequential event propagation in graph, such as retweeting by social network users, news transmitting between websites, etc. Given a collection of event propagation sequences, conventional point process model consider only the event history, i.e. embed event history into a vector, not the latent graph structure.
	We propose a Graph Biased Temporal Point Process (GBTPP) leveraging the structural information from graph representation learning, where the direct influence between nodes and indirect influence from event history is modeled respectively. Moreover, the learned node embedding vector is also integrated into the embedded event history as side information. Experiments on a synthetic dataset and two real-world datasets show the efficacy of our model compared to conventional methods and state-of-the-art.
\end{abstract}


\IEEEpeerreviewmaketitle

\section{Introduction}

Event sequences modeling is widely used across different areas and applications. In e-commerce, the on-line purchase records over time can be modeled as event sequences. In health informatics, series of treatments taken by patients can be tracked as event sequences. In seismology, a sequence of earthquakes recorded are modeled as event sequences. In social media like Twitter, every time a user posts, transmits or likes a tweet, it corresponds to a new event adding to the user behavior sequence. In all the above settings, event sequences modeling is of vital importance for predicting future events and recognizing hidden patterns given history sequences.

For modeling event sequences, Temporal Point Processes (TPP)~\cite{daley2007introduction} is a useful tool.
For example, \cite{ZhouAISTATS13} uses the so-called multi-dimensional Hawkes processes to model the sequential user actions in a social network, and the learned infectivity matrix is useful for uncovering the mutual influences between users. Mixtures of Hawkes processes \cite{li2014learning} are modeled for inferring missing event attributes from the behavioral observation by considering the dependency among dyadic events.
In \cite{yan2013towards}, a water pipe failure prediction system is designed for effective replacement and rehabilitation. The water pipe failure sequence is formulated as a self-exciting stochastic process.

Marked Temporal Point Process is (MTPP) an important domain in TPP for event sequences modeling. In MTPP, an event can carry extra information called marker. The marker typically refers to event type and lies in the discrete label space, i.e. a finite category set $\{1, ..., m\}$. In e-commerce, the marker can refer to the users and items. In health informatics, markers can be the treatments and medications of a patient. In predictive maintenance, markers can carry important log data for when the failure occurs and what is the type. In all these examples, effectively modeling and predicting the dynamic behavior while leveraging the information contained in the markers is of vital importance for MTPP. 

In this paper, we focus on a special case of MTPP, where the event sequence is an event propagation process in a directed weighted graph and the marker denote the node in the graph. For example, a retweeting sequence in social network where the markers denote users in user network, a news transmission sequence between websites where the markers denote the website in the influence network.

\begin{figure}[tb!]
	\centering
	\subfigure{\includegraphics[width=0.40\textwidth]{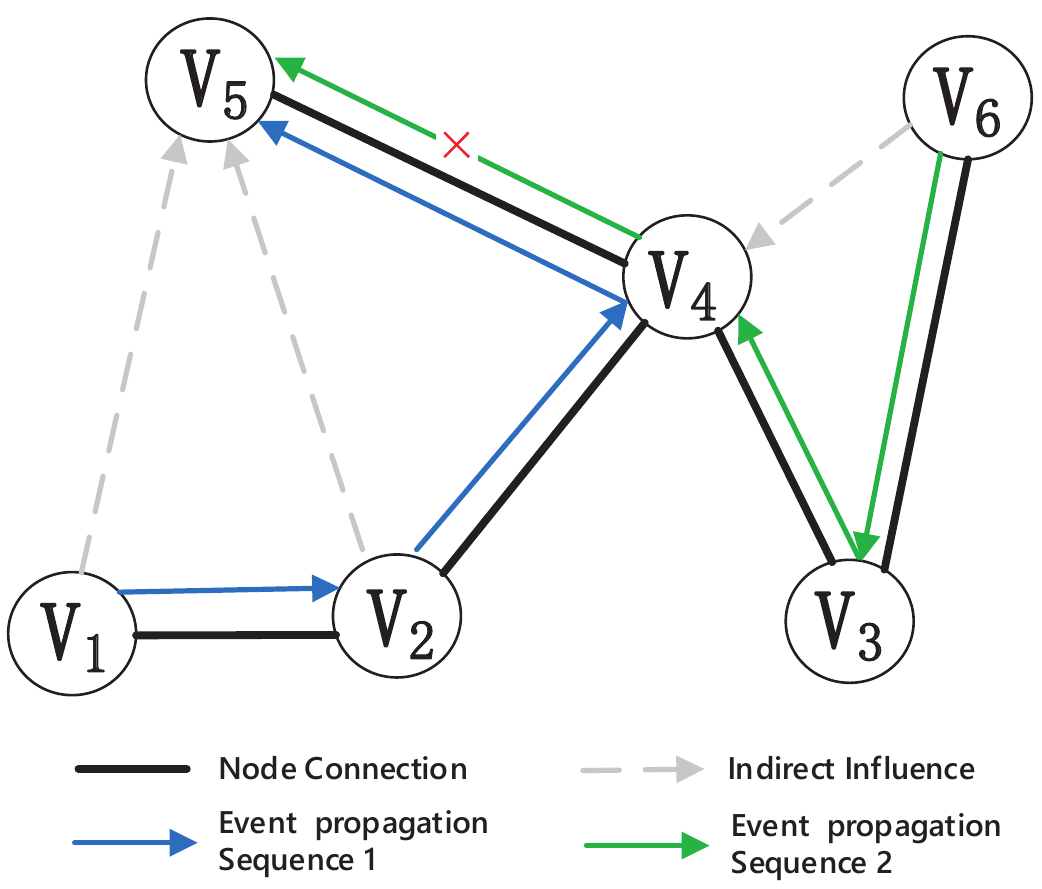}}
	\caption{An example of event propagation in a sample graph with two observed event propagation sequences. We want to model the event propagation process given observed two propagation sequences $\{V_1\rightarrow V_2 \rightarrow V_4\rightarrow V_5\}$ and $\{V_6 \rightarrow V_3 \rightarrow V_4\}$. Considering only the node connections can not cope with the situation that only events from $V_2$ propagate to $V_5$ through $V_4$ but events from $V_3$ can not propagate to $V_5$. While conventional Temporal Point Process (TPP) can deal with this case, it measures the indirect influence from node $V_1$, $V_2$ to $V_5$ and $V_6$ to $V_4$ equally as direct influence between connected nodes $V_4$, $V_5$ and $V_3$, $V_4$ which is also inaccurate.}
	\label{fig:example}
	\vspace{-10pt}
\end{figure}

To model and predict event propagation is a challenging task. The difficulty lies in how to leverage the network structure and node proximity in the graph when modeling the event propagation sequence.
Conventional TPP methods model the event propagation path as general event sequences, computing probabilities and making predictions basing on history events, like in \cite{du2016recurrent,yan2013towards,XiaoAAAI17,Wu:2018:DLF:3219819.3220035}. But modeling event propagation without considering the connections of nodes in graph is inaccurate. As shown in Fig.~\ref{fig:example}, in conventional TPP model, node $V_1$ or $V_2$ is not connected to node $V_5$ while the indirect history influence is measured equally as the direct influence between connected node $V_4$ and $V_5$.

Intuitively, only the connected nodes have influence on each other. However, when we predict the propagation of events in graph, merely considering the direct influence between connected nodes is also not appropriate without using TPP method. As in Fig.~\ref{fig:example}, for the case that only events from $V_2$ propagate to $V_5$ through $V_4$, while events from $V_3$ can not propagate to $V_5$, measuring only the direct influence can not handle this situation.

In this paper, to model the event propagation process in graph considering both direct influence between connected nodes and indirect influence of propagation history, we propose a Graph Biased Temporal Point Process (GBTPP) leveraging the structural information extracted from graph representation. Compared with conventional TPP model, we make two major contributions:

i) The direct influence between connected nodes is measured separately from the indirect history influence as a bias term, leveraging the first-order proximity between nodes learned by node embedding. The intensity of the direct influence is controlled by a scale factor related to the event history.

ii) The node embedding vector is added to the event propagation history embedding when modeling the indirect influence, so that the structural information can be integrated into the model.

To verify the efficacy of our model, we experiment on a synthetic dataset and two real-world datasets, including a Higgs Twitter Dataset\cite{de2013anatomy} predicting tweets propagation in social network Twitter, and a MemeTracker Dataset\cite{Leskovec:2009:MDN:1557019.1557077} predicting meme propagation between websites. Empirical results show that the proposed GBTPP model outperforms conventional methods and state-of-the-art one.

\section{Related Work}

Sequential event data is generated from lots of social activities, e.g. financial transactions, electronic health records, e-commerce purchase records, etc. In these scenarios, the sequential event data contains abundant information about which type of event happens at what time. 
For example, the daily routine of a person contains various places at different moments during one day. Stock managers buy or sell stocks at different instants of time. Patients with chronic diseases pay regular visits to the hospital to obtain their diagnoses each time. 

Moreover, there are many underlying structures in sequential data besides its intrinsic temporal structure, especially for network and graph structure. Actually, when modeling 

\subsection{Sequential Data Modeling}

As sequential event data is frequently produced from various domains and applications, modeling the event sequences, especially predicting future events is of vital importance: based on the observed event sequence history, predicting which type of event will happen at what time in the future. 
This kind of prediction task is of great use in many applications, e.g. in stock market, predicting when to buy or sell a particular stock has important business value. For mainstream assistant, making spatial and temporal predictions on when and where a person will visit a certain place will make personal service more suitable and relevant. For health-care services, predicting future clinical events and disease progression can help to provide personal medical services and reduce potential risks. 

To model event sequences, existing literatures attempt to solve this problem in mainly two categories of methods:

First, the conventional varying-order Markov models\cite{begleiter2004prediction} deal with this problem as a discrete-time sequence prediction task. 
Based on the observed history states sequence, prediction of the event type is given by the most likely state that the state transition process will evolve into on the next step. 
An obvious limit for the families of Markov models is that they assume the state transition process proceed with unit time-step, it can not capture the temporal dependency of the continuous time and give predictions on the exact time of the next event. 
Moreover, Markov models can not deal with long dependency of the history events when the event sequence is long, because the size of the state space grow exponentially with the number of the time steps considered in Markov model. 
It is worth mentioning that semi-Markov models \cite{janssen2013semi} can model continuous time-intervals between two states to some extent, by assuming the intervals to follow some simple distributions, but it still has the state space explosion problem when dealing with long time dependency. 

Second, Temporal point processes with conditional intensity functions is a more general framework for sequential event data modeling.  
Temporal Point Process (TPP) is powerful for modeling event sequence with time-stamp in continuous time space. Early work dates back to the Hawkes processes \cite{HawkesBiometrika71} which shows appropriateness for self-exciting and mutual-exciting process like earthquake and its aftershock \cite{ogata1988statistical, ogata1998space}. As an effective model for event sequence modeling, TPP has widely used in various applications, including data mining tasks e.g. social infectivity learning \cite{li2013dyadic},
conflict analysis \cite{MangionPNAS12},
crime modeling \cite{StomakhinIP11},
email network analytics \cite{FoxJASA16} and extremal behavior of stock price \cite{EmbrechtsJAP11}, and event prediction tasks e.g. failure prediction \cite{ErtekinRPP2015}, sales outcome forecasting \cite{YanAAAI15}, literature citation prediction \cite{wang2013quantifying}.

Traditional TPP models are modeled by parametric forms involving manual design of conditional intensity function $\lambda(t)$ depicting event occurrence rate over time, which measures the instantaneous event occurrence rate at time $t$. A few popular examples include:
\begin{itemize}
	\item Poisson process\cite{kingman1993poisson}: the basic form is history independent $\lambda(t)=\lambda_{0}$ which can be dated back to the 1900's;
	\item Reinforced Poisson processes \cite{PemantlePS07}: the model captures the `rich-get-richer' mechanism by $\lambda(t)=\lambda_0f(t)i(t)$ where $f(t)$ mimics the aging effect while $i(t)$ is the accumulation of history events;
	\item Self-exciting process (Hawkes process) \cite{hawkes1971point}: it provides an additive model to capture the self-exciting effect from history events $\lambda(t)=\lambda_0+\sum_{t_i<t}g_{exc}(t-t_i)$;
	\item Reactive point process \cite{ErtekinRPP2015}: generalization to the Hawkes process by adding a self-inhibiting term to account for the inhibiting effects from history $\lambda(t)=\lambda_0+\sum_{t_i<t}g_{exc}(t-t_i)-\sum_{t_i<t}g_{inh}(t-t_i)$.
\end{itemize}

One obvious limitation of the above TPP models is that they all assume all the samples obey a single parametric form which is too idealistic for real-world data. By contrast, recurrent neural network (RNN) based models \cite{du2016recurrent,mei2017neural,XiaoAAAI17} are devised for learning point process.
In these works, recurrent neural network (RNN) and its variants e.g. long-short term memory (LSTM) are used for modeling the conditional intensity function over time. More recently attention mechanisms are introduced to improve the interpretability of the neural model \cite{wang2017cascade}. 

\subsection{Sequential Data modeling with Graph Structure}
With the increase of data dimensionality and the data types become more diverse, sequential data often contains other structures, typically graphs for example. 
The most common type of graph data associated with sequential data is the spatial graph structure, which is in the form of spatio-temporal data. The spatio-temporal point process is often used to describe the spatial and temporal structure of spatio-temporal data. 
For example, in \cite{etherington2012using}, the spatio-temporal point process is used to establish the spatio-temporal grain of continuous landscape tessellations and graphs, defining the the grain of landscape tessellations and graphs with a variety of geometries. One of the cases is that, the spatio-temporal data describing the spatial activity of brushtail possums is collected using very high frequency (VHF) and global positioning system (GPS) based radiotelemetry, given their associated temporal scale. Then the intensity of spatio-temporal point process is computed basing on the spatio-temporal data, and the grain of regular or irregular landscape tessellations or graphs in terms of point process intensity provides a general way of reporting grain.

Most recently, the spatio-temporal point process has also been used to capture crime linkage in \cite{zhu2019crime}. Actually, a spatio-temporal-textual point process is proposed to utilize the text, time and location information in the crime records. Specifically, in the conditional intensity function, the temporal effect is measured by the temporal kernel function, e.g. exponential function; the spatial correlation is measured by a coefficient matrix $A=\{\alpha_{ij}\}$ where each entry $\alpha_{ij}$ represents the strength of the influence from location indexed $j$ by to the location indexed by $i$; the text information is represented by a mapping function to project the bag-of-words representations into a $m$-bit binary Modus Operandi embedding space, where the embedding similarity is measured by inner product. 

In addition to spatio-temporal point processes dealing with spatial graph information for sequential data, from the perspective of graph representation, exploring the influence of sequential data on graphs is another important issue, e.g. learning and representing dynamic graphs. In most recent work in \cite{singer2019node}, an end-to-end architecture leveraging Recurrent Neural Networks (RNNs) is proposed to create a temporal node embedding over dynamic graphs. Specifically, by introducing the RNNs to reduce the sequential data into a temporal historical embedding, the model is able to calculate the final node embedding combining the temporal dynamics with the graph structures. Compared with conventional works on dynamic graphs designed for specific prediction tasks, the node embedding method over dynamic graphs in \cite{singer2019node} is a more generalized graph embedding technique that jointly optimize the node representations and prediction task objectives like node classification and link prediction. 

As discussed above, from the view of application, we consider existing literatures \cite{etherington2012using, singer2019node, zhu2019crime} to be related to our work, as they all leverage information from graph structures, considering both temporal dynamic information and structural information. 
While from the view of methodology, there are differences and connections between our work and these existing ones. Especially, they encode the structural information into the temporal point processes in different ways.

Specifically, the spatio-temporal point process model implemented in \cite{etherington2012using, zhu2019crime} encode the spatial graph information by adding an extra marker dimension in conventional parametric Marked Temporal Point Processes (MTPP).  
Conventional MTPP uses conditional intensity function $\lambda\left(t, x | \mathcal{H}_{t}\right)=\mu+\sum_{j: t_{j}<t} g\left(t, t_{j}, x, x_{j}\right)$, where $x$ is the marker representing event type, $\mu$ is the constant base intensity, $\sum_{j: t_{j}<t} g\left(t, t_{j}, x, x_{j}\right)$ is the cumulative mutual intensity of event history $\mathcal{H}_{t} = \{x_{j}, t_{j}\}_{j: t_{j}<t}$ and $g$ is the nonnegative triggering function\cite{reinhart2018review}.
In \cite{zhu2019crime} where the spatio-temporal point process is used for crime linkage detection, an extra marker $s$ is introduced representing crime location in the spatial graph, so that the spatial graph information can be encoded into the intensity function by $\lambda\left(s, t, x | \mathcal{H}_{t}\right)=\mu(s)+\sum_{j: t_{j}<t} g\left(s, s_{j}, t, t_{j}, x, x_{j}\right)$. 
In our work, instead of building a conventional parametric model as in \cite{zhu2019crime}, we adopt Recurrent Neural Network (RNN) based models and incorporate the structural information via graph embedding technique. 

While recent work in \cite{singer2019node} also employs the combination of RNN and graph embedding, it is in essence different from our work. In short, \cite{singer2019node} focus on graph embedding learning over temporal dynamic graph and modifies existing graph embedding technique by introducing RNN, while we focus on the information propagation process between nodes in a static graph and revise existing RNN based MTPP model by introducing graph embedding.
In specific, there are two major differences methodologically compared to our work: i) For the use of RNN, \cite{singer2019node} simply use RNN to reduce each node's historical embeddings $X^{(v)}$ to a $d$-dimension node embedding vector, where $X^{(v)} \in \mathbb{R}^{T \times d}$ is the historical $T$ embeddings of node $v$ over time and each of size $d$. Instead of simply extracting temporal dynamic information from sequential data, we employ RNN as a component of the MTPP model, via encoding the information propagation history into a hidden vector $\boldsymbol{h}_n$, and further use $\boldsymbol{h}_n$ for learning the the parameters of MTPP's intensity function. 
ii) For the use of graph embedding, \cite{singer2019node} uses static graph embedding as initialization of the algorithm and the final node embedding vector is obtained by reducing the historical $T$ embeddings of each node into a final one via RNN, so that the evolution of a temporal graph over time can be captured. While in our work, the vanilla graph embedding is used to acquire each node's embedding vector, as we focus on the propagation process in static graph. 

Though sequential data with spatial graph information is explored in spatio-temporal point processes \cite{etherington2012using, zhu2019crime}, when dealing with more general graph structure e.g. information propagation in social networks, conventional spatio-temporal point processes are not able to incorporate the graph structural information. 
While in \cite{singer2019node}, a graph embedding technique is developed over temporal dynamic graphs for graph representation rather than sequential data modeling, temporal point processes modeling on sequential data with graph structure is rarely studied. 

In this paper, we study this problem in the form of event propagation in social networks. Event or information propagation and diffusion in social networks has been studied under different methodologies including TPP. For example, in \cite{karmaker2017modeling}, Hawkes process is used to model the evolutionary trend of the influence of events on search queries by proposing computational measures that quantify the influence of an event on a query to identify triggered queries.
When dealing with event propagation sequences, a major limitation of existing studies is that the structural information of the latent graph $\mathcal{G}=(V, E)$ is not utilized. Conventional TPP models including state-of-the-art method in \cite{du2016recurrent} solve event propagation modeling as general event sequences modeling and take input $\{v_i, t_i\}$, while our GBTPP model leverage the structural information and node proximity of graph $\mathcal{G}$ taking input $\{v_i,t_i,\boldsymbol{y}_i\}$, where $\boldsymbol{y}_i$ is the node embedding vector obtained by a graph representation learning method for $\mathcal{G}$.

\section{Proposed Model}

Given a collection of observed event propagation sequences $\mathcal{C}=\left\{\mathcal{S}^{1}, \mathcal{S}^{2}, \ldots\right\}$ in a latent graph $\mathcal{G}=(V, E)$ with node set $V$ and edges set $E$, each $\mathcal{S}^{k}=\left(\left(t_{1}^{k}, v_{1}^{k}\right),\left(t_{2}^{k}, v_{2}^{k}\right), \ldots\right)$ is a sequence of pairs $\left(t_{j}^{k}, v_{j}^{k}\right)$, where $t_j^k$ is the time when event $k$ propagate from node $v_j$ to node $v_{j+1}$.

To model the event propagation processes and make predictions on propagation node $v_{n+1}$ and propagation time $t_n$ while leveraging the structural information in graph $\mathcal{G}$, our method contains two steps:
\begin{itemize}
	\item \textbf{Graph Representation} Learn a representation for latent graph $\mathcal{G}=(V, E)$ from observed propagation sequences $\mathcal{C}$. For each node $v_i\in V$, we learn the node embedding vector $\boldsymbol{y}_i$ that preserve the first order proximity.
	\item \textbf{Graph Biased Temporal Point Process} Train a GBTPP model based on the learned graph representation $\{\boldsymbol{y}_{k}\}_{k=1}^{V}$ and observed propagation sequences $\mathcal{C}$. The GBTPP model integrate the node proximity as a bias term, and use a scale factor to control intensity of this term.
\end{itemize}

We present the details of the \emph{Graph Representation} and \emph{Graph Biased Temporal Point Process} (GBTPP) as follows.

\subsection{Graph Representation}
Graph embedding and representation has been widely used in both academia and industry in recent years. Lots of literatures are proposed to convert a graph $\mathcal{G}=(V, E)$ into a $d$-dimensional space, in which the graph property is preserved\cite{cai2018comprehensive}. In graph representation techniques, the graph is represented as either a d-dimensional vector (for a whole graph) or a set of d-dimensional vectors with each vector representing the embedding of part of the graph (e.g., node, edge, substructure).

In general, the graph property is quantified by proximity measured by the \emph{first-order proximity} and \emph{second-order proximity}:
\begin{itemize}
	\item \emph{First-order proximity}: The local pairwise similarity between nodes connected by edges. It
	compares the direct connection strength between a node pair. The first-order proximity between node $v_i$ and node $v_j$ is the weight of the edge $e_{ij}$, i.e., $A_{i,j}$. Two nodes are more similar if they are connected by an edge with larger weight.
	\item \emph{Second-order proximity}: The similarity of the nodes' neighborhood structures. The more similar two nodes' neighborhoods are, the larger the second-order proximity value between them. Formally, the second-order proximity $s^{(2)}_{ij}$ between node $v_i$ and $v_j$ is a similarity between $v_i$'s neighborhood $s^{(1)}_i$ and $v_j$'s neighborhood $s^{(1)}_j$.
\end{itemize}
In general, the learned graph representation preserve either \emph{first-order proximity} like \cite{man2016predict,wei2017cross}, or \emph{second-order proximity} like \cite{zhang2017regions,liu2016aligning,feng2016gake}, etc. In some recent work\cite{DBLP:conf/www/TangQWZYM15,DBLP:conf/emnlp/BordesCW14, DBLP:conf/sigir/ZhangW16}, both the first-order and second-order proximities are empirically calculated based on the joint probability and conditional probability of two nodes.

In this paper, the \emph{first-order proximity} is preserved. We learn the graph representation as a set of embedding vectors for the nodes in latent graph $\mathcal{G}=(V, E)$ keeping the \emph{first-order proximity}, where each node $v$ in graph $\mathcal{G}$ is represented by two $d$-dimensional vector $\boldsymbol{y}^s$ and $\boldsymbol{y}^e$.
For the directed weighted graph $\mathcal{G}$, the direct influence from node $v_i$ to node $v_j$ i.e. the weight of edge $e_{ij}$ is computed as $p(\boldsymbol{y}_i^s, \boldsymbol{y}_j^e)$, and the direct influence from $v_j$ to $v_i$ is computed as $p(\boldsymbol{y}_j^s, \boldsymbol{y}_i^e)$.

Specifically, to obtain the node embedding for each node in $\mathcal{G}=(V, E)$, we have the following implementation.

\subsubsection{Edge Reconstruction Probability}
The learned node embedding should be able to re-establish edges in the original input graph. This can be realized by maximizing the probability of generating all observed edges
using node embedding. The directed edge between a node pair $v_i$ and $v_j$ indicating their first-order proximity can be calculated as the joint probability using the embedding $\boldsymbol{y}_i^s$ of $v_i$ and $\boldsymbol{y}_j^e$ of $v_j$:
\begin{equation}\label{eq:1st_order}
p^{(1)}\left(v_{i}, v_{j}\right)=\frac{1}{1+\exp \left(-{\boldsymbol{y}_{i}^{s}}^T \boldsymbol{y}_{j}^e\right)}.
\end{equation}

\subsubsection{Minimizing Distance-based Loss}
From event propagation sequences $\mathcal{C}$ in weighted directed graph $\mathcal{G}$, we have the empirical estimation of the adjacent matrix $A$, in which $A_{i,j}$ is the empirical estimation for the weight of edge $e_{ij}$ computed by the normalized propagation number from $v_i$ to $v_j$ as $A_{i,j}=\frac{N_{ij}}{N_{max}}$, where $N_{ij}$ is the number of observed event propagation from $v_i$ to $v_j$ and $N_{max}$ is the global maximum number of event propagation between any given node.

To capture the structural information and connections between nodes in graph $\mathcal{G}$, the node proximity calculated based on node embedding in Eq.\ref{eq:1st_order} should be as close to the node proximity calculated based on the observed edges as possible.
Specifically, node proximity can be calculated based on node embedding or empirically calculated based on observed edges. Minimizing the differences between the two types of proximities can preserve the corresponding proximity.

For the \emph{first-order proximity}, it can be computed as $p^{(1)}$ using node embedding defined in Eq.\ref{eq:1st_order}, while the empirical probability is
$
\hat{p}^{(1)}\left(v_{i}, v_{j}\right)=A_{i,j} / \sum_{e_{i j} \in E} A_{i,j}
$,
where $A_{i,j}$ is the empirical estimation for the weight of edge $e_{ij}$. The smaller the distance between $p^{(1)}$ and $\hat{p}^{(1)}$ is, the better first-order proximity is preserved.

By adopting KL-divergence as the distance function, we can minimize the difference between $p^{(1)}$ and $\hat{p}^{(1)}$, and the objective function preserving the first-order proximity in :
\begin{equation}\label{eq:}
\mathcal{O}_{\min }^{(1)}=\min -\sum_{e_{i j} \in E} A_{i j} \log p^{(1)}\left(v_{i}, v_{j}\right).
\end{equation}
For each node $v_i$, we can learn the corresponding node embedding vector $\boldsymbol{y}_i=\{\boldsymbol{y}_i^s, \boldsymbol{y}_i^e\}$ by Eq.\ref{eq:}, indicating the \emph{first order proximity} by Eq.\ref{eq:1st_order}.

\subsection{Graph Biased Temporal Point Process}
\begin{figure}[tb!]
	\centering
	\subfigure{\includegraphics[width=0.48\textwidth]{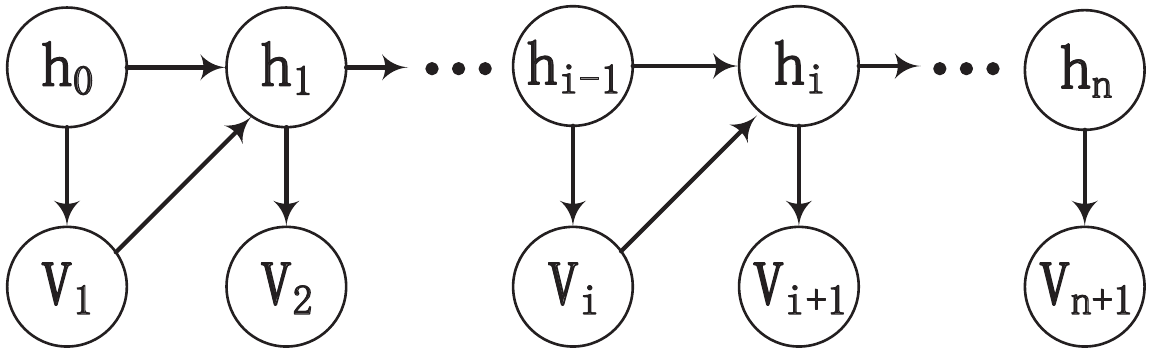}}
	\vspace{-5pt}	
	\caption{Conventional Temporal Point Process (TPP): Current node $v_i$ is embedded into history vector $\boldsymbol{h}_i$, and the prediction of the next propagation node $v_{n+1}$ is given by $P(v_{n+1}|\boldsymbol{h}_n)$.}
	\label{fig:alg_conv}
\end{figure}
\begin{figure}[tb!]
	\centering
	\subfigure{\includegraphics[width=0.42\textwidth]{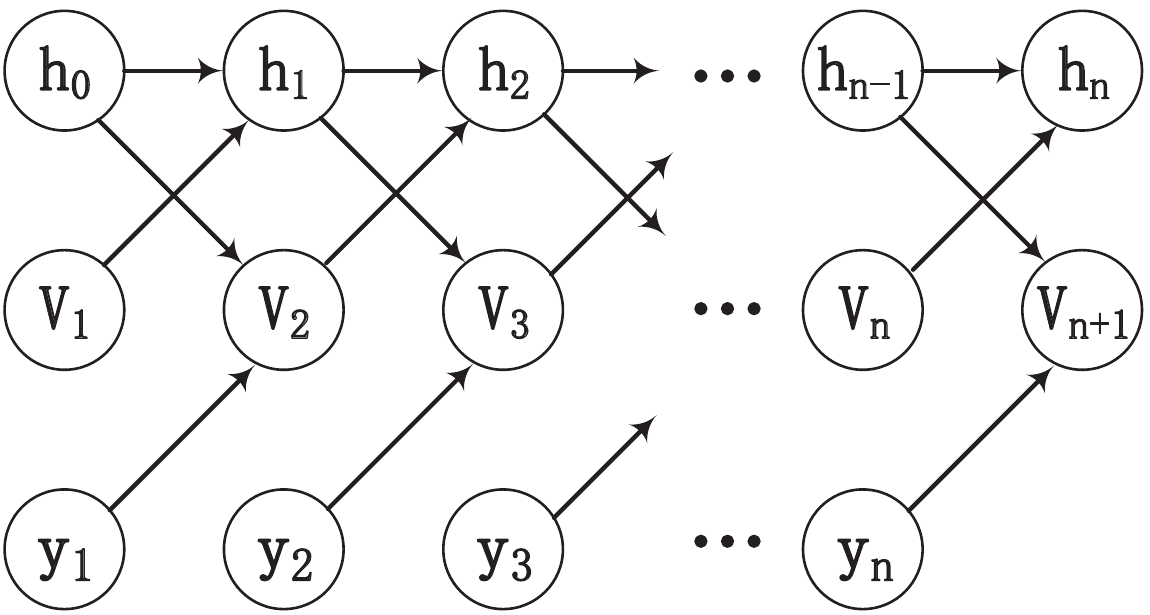}}
	\vspace{-5pt}	
	\caption{Graph Biased Temporal Point Process (GBTPP): The direct influence from current node $v_n$ is measured separately from propagation history $\boldsymbol{h}_{n-1}$ by the node embedding vector $\boldsymbol{y}_n$. The prediction of the next propagation node $v_{n+1}$ is given by $P(v_{n+1}|\boldsymbol{h}_{n-1}, \boldsymbol{y}_n)$.}
	\label{fig:alg_new}
	\vspace{-0pt}
\end{figure}
\begin{figure}[tb!]
	\centering
	\subfigure{\includegraphics[width=0.48\textwidth]{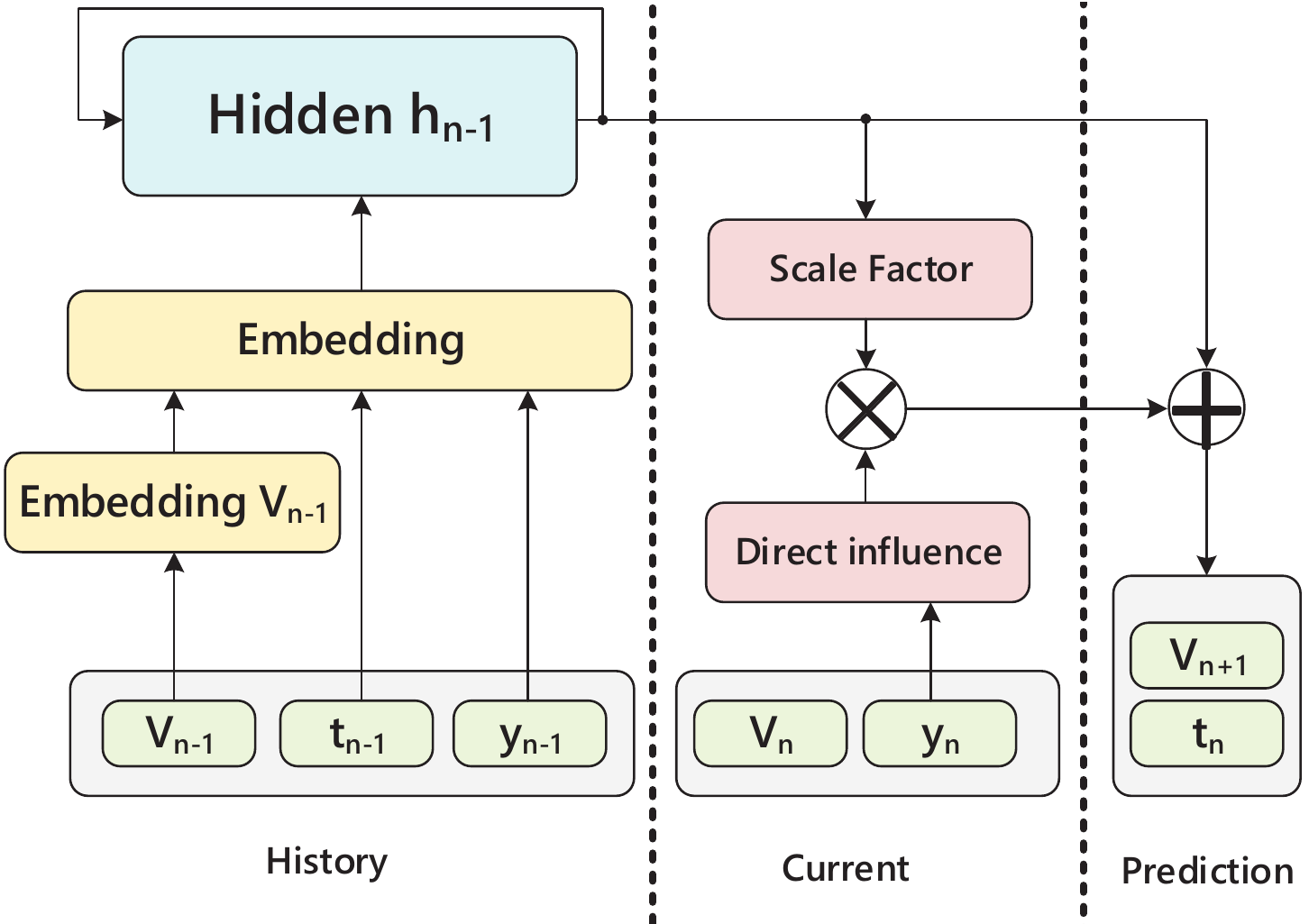}}
	\vspace{-5pt}	
	\caption{Architecture for Graph Biased Temporal Point Process (GBTPP). The event propagation history $\mathcal{H} = \{(v_1,t_1,\boldsymbol{y}_1), (v_2,t_2,\boldsymbol{y}_2),\dots,(v_{n-1},t_{n-1}, \boldsymbol{y}_{n-1})\}$ is embedded recurrently into $\boldsymbol{h}_{n-1}$ by an input embedding layer and a recurrent layer. The direct influence of the current node $v_n$ is computed by the proximity between nodes using node embedding $\boldsymbol{y}_n$, with a scale factor controlling the intensity of the influence computed by $\boldsymbol{h}_{n-1}$.}
	\label{fig:alg}
	\vspace{-20pt}
\end{figure}
Given history propagation sequence $\mathcal{H}_{n-1}=\big((t_{1}, v_{1}),(t_{2}, v_{2})$ $ \dots, (t_{n-1}, v_{n-1})\big)$ and current node $v_n$, the Graph Biased Temporal Point Process (GBTPP) model aims to compute the probability $P(v_{n+1}|\mathcal{H}_{n-1},v_{n})$ of the event propagating to node $v_{n+1}$ given propagation history $\mathcal{H}_{n-1}$ and current node $v_n$, and the estimation of the propagation time $t_n$ by the likelihood $f(t_n|\mathcal{H}_{n-1},v_{n})$.

\begin{figure}[tb!]
	\centering
	\subfigure{\includegraphics[width=0.48\textwidth]{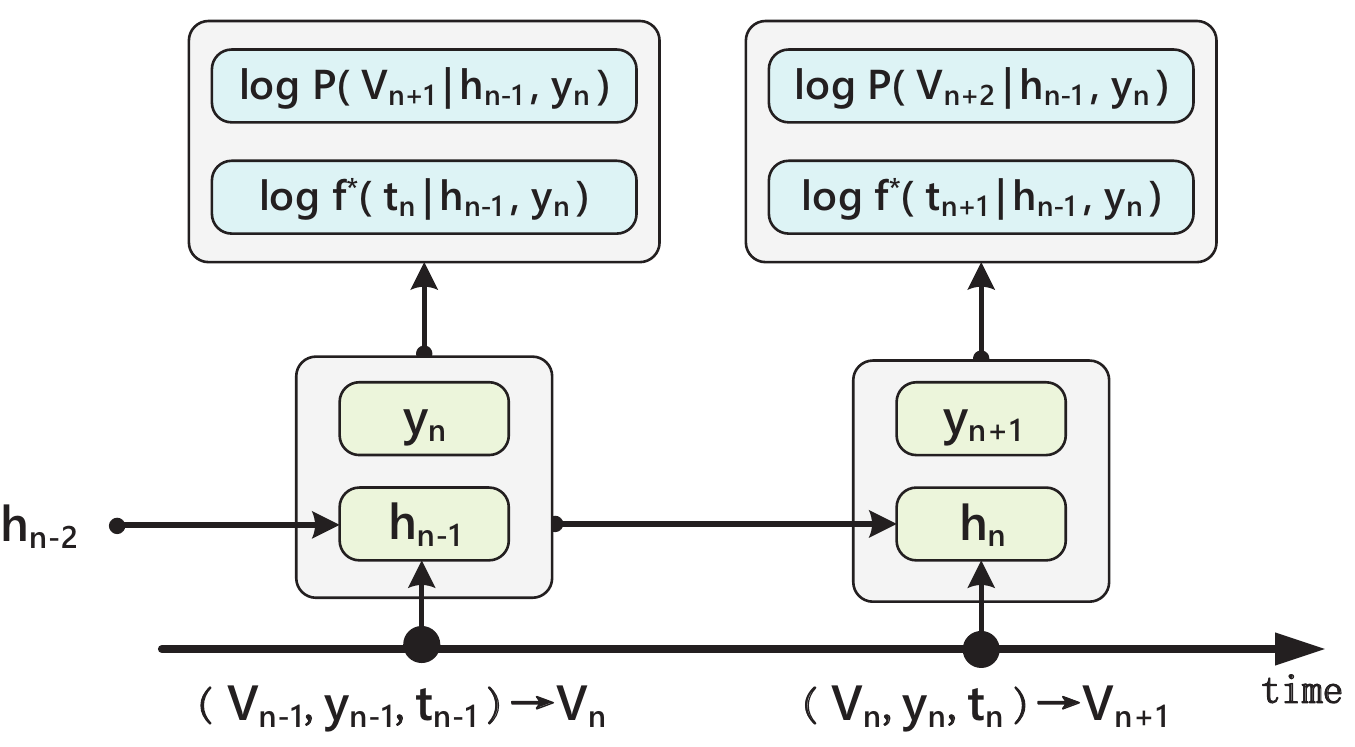}}
	\vspace{-5pt}	
	\caption{Illustration for the Graph Biased Temporal Point Process (GBTPP) model. We have the observation that an event propagate from node $v_{n-1}$ to $v_n$ at time $t_{n-1}$. The GBTPP aims to make predictions on the next propagation node $\hat{v}_{n+1}$ and next propagation time $\hat{t}_n$, i.e. predicting that the event will propagate from node $v_{n}$ to $\hat{v}_{n+1}$ at time $\hat{t}_n$. To make this happen, we embed the last propagation history embedding $\boldsymbol{h}_{n-2}$ and features of last history node $\{v_{n-1},\boldsymbol{y}_{n-1},t_{n-1}\}$ into new history embedding $\boldsymbol{h}_{n-1}$ as Eq.\ref{eq:history}, together with the node embedding vector $\boldsymbol{y}_{n}$ of current node $v_n$, we compute the log-likelihood of the propagation node and time by Eq.\ref{eq:P} and Eq.\ref{eq:ll_t} respectively. The node embedding vector $y_{n-1}$ and $y_n$ is pre-learned by graph representation. }
	\label{fig:illustrate}
	\vspace{-10pt}
\end{figure}

As shown in Fig.~\ref{fig:alg_conv} and Fig.~\ref{fig:alg_new}, conventional Temporal Point Process (TPP) model embeds current node $v_n$ into history embedding vector $\boldsymbol{h}_n$, while the GBTPP model measures the direct influence of current node by $\boldsymbol{y}_n$ and indirect history influence by $\boldsymbol{h}_{n-1}$ respectively. The architecture of GBTPP is presented in Figure.\ref{fig:alg}. We also illustrate GBTPP model in Fig.~\ref{fig:illustrate}. Specifically, we specify the proposed model as the following parts: \emph{Input Embedding}, \emph{History Embedding}, \emph{Graph Bias} term computation and \emph{Prediction}.

\subsubsection{Input Embedding}
As in Fig.~\ref{fig:alg}, the history input includes $\{v_{n-1}, t_{n-1}, \boldsymbol{y}_{n-1}\}$ as a triple, including a sparse one-hot vector $v_{n-1}$ representing a node in graph $\mathcal{G}=(V, E)$, a continuous value $t_{n-1}\in (0,T]$ indicating the time of the event propagate from node $v_{n-1}$ to $v_n$, and the corresponding node embedding vector $\boldsymbol{y}_{n-1}\in \mathbb{R}^d$.

The sparse one-hot vector representation of the node $v_i$ is projected into a latent space by an embedding layer with the weight matrix $\boldsymbol{W}_{e m}$ to achieve a more compact and efficient representation as $\boldsymbol{v}_{i}=\boldsymbol{W}_{em}^{\top} v_{i}+\boldsymbol{b}_{em}$, where $\boldsymbol{v}_{i}$ is the embedding for $v_i$. Then the representation vector $\boldsymbol{v}_{i}$ is embedded into a common feature space $\mathbb{R}^H$ for both input embedding and next step history embedding, with a weight matrix $\boldsymbol{W}^v$.

For the propagation time input $t_{n-1}$ , we can extract the associated temporal features, e.g. the inter-event duration $d_{n-1} = t_{n-1} - t_{n-2}$. Here we slightly abuse the notation for temporal feature still as $t_{n-1}$ for conciseness. The temporal feature $t_{n-1}$ is also embedded into common feature space $\mathbb{R}^H$ a by weight matrix $\boldsymbol{W}^t$.

Similarly, the node embedding $\boldsymbol{y}_i$ is also projected from the node embedding space to the feature space by an embedding layer with weight matrix $\boldsymbol{W}^y$. The history input triple $\{v_{n-1}, t_{n-1}, \boldsymbol{y}_{n-1}\}$ is embedded into a common history feature space as $\{\boldsymbol{W}^{v} \boldsymbol{v}_{n-1} ,\boldsymbol{W}^{y} \boldsymbol{y}_{n-1},\boldsymbol{W}^{t}t_{n-1}\}$.

\subsubsection{History Embedding}
In History Embedding part, the embedded input
is added to the propagation history embedding vector $\boldsymbol{h}_{n-1}$ with the last propagation trajectory embedding vector $\boldsymbol{h}_{n-2}$ by a Recurrent Neural Network, so that we have an event propagation history embedding $\boldsymbol{h}_{n-1}$ as
\begin{align}\notag\label{eq:history}
\boldsymbol{h}_{n-1}=\max \Bigg\{ &
\boldsymbol{W}^{v} \boldsymbol{v}_{n-1} + \boldsymbol{W}^{y} \boldsymbol{y}_{n-1} + \boldsymbol{W}^{t}t_{n-1} \\ & + \boldsymbol{W}^{h} \boldsymbol{h}_{n-2}+\boldsymbol{b}_{h},
0\Bigg\}.
\end{align}
Compared with conventional temporal point processes, except for event marker i.e. propagation node $v_{n-1}$ and time $t_{n-1}$, the node embedding $\boldsymbol{y}_{n-1}$ indicating the structural information of $v_{n-1}$ in graph $\mathcal{G}=(V, E)$ is used as side information input when computing history embedding.

\subsubsection{Graph Bias}
Given the embedded event propagation history, conventional temporal point processes compute the event propagation probability as $P(v_{n+1}|\mathcal{H}_n)$ and the likelihood of time $t_n$ as $f(t_n|\mathcal{H}_n)$. For example, in \cite{du2016recurrent}, the propagation probability is computed as
\begin{equation}\label{eq:conv_P}
P\left(v_{n+1}=k | \boldsymbol{h}_{n}\right)=
\frac{\exp \left(\boldsymbol{V}_{k, :}^{h} \boldsymbol{h}_{n}+b_{k}^{h}\right)}{\sum_{k=1}^{V} \exp \left(\boldsymbol{V}_{k, :}^{h} \boldsymbol{h}_{n}+b_{k}^{h}\right)},
\end{equation}
where $V$ is the number of nodes, $\boldsymbol{V}_{k, *}^{h}$ is the $k$-th row of parameter matrix $\boldsymbol{V}$, and $b_k^h$ is the constant bias term. The conditional intensity function $\lambda(t)$ is also computed conditional on $h_n$ by
\begin{equation}\label{eq:conv_lambda}
\lambda(t)=\exp \left(\boldsymbol{v}^{t \top} \cdot \boldsymbol{h}_{n}+w^{t}\left(t-t_{n-1}\right)+b^{t}\right),
\end{equation}
where $\boldsymbol{v}^{t}$ is a column vector and $w^t$, $b^t$ is scalar, and the likelihood of event propagation time $t_n$ is  computed as $f(t_n)=\lambda(t) \exp \left(-\int_{t_{n-1}}^{t} \lambda(\tau) d \tau\right)$.

Compared with conventional methods that embed current node $v_n$ into propagation history $\boldsymbol{h}_{n}$, we model the direct influence of current node and the indirect influence of the propagation history respectively. Besides using a constant bias term as $b_k^h$ as in Eq.\ref{eq:conv_P}, a graph bias term $b(\boldsymbol{h}_n, \boldsymbol{y}_n, \boldsymbol{y}_k)$ is introduced for event propagation propagation probability as:
\begin{equation}\label{eq:bias}
b(\boldsymbol{h}_{n-1},\boldsymbol{y}_n, \boldsymbol{y}_k) = \text{ReLU}(\boldsymbol{U}_{n,:}^{h} \boldsymbol{h}_{n-1})p(\boldsymbol{y}_n, \boldsymbol{y}_k),
\end{equation}
where $p(\boldsymbol{y}_n, \boldsymbol{y}_k)$ is the \emph{first-order proximity} learned in the graph representation step measuring the direct influence of node $v_n$ to node $v_k$ as in Eq.\ref{eq:1st_order}, and $\text{ReLU}$ function $\text{ReLU}(\boldsymbol{U}_{n,:}^{h} \boldsymbol{h}_{n-1})$ compute the scale factor that measures the intensity of this influence, $\boldsymbol{U}_{n, :}^{h}$ is the $n$-th row of parameter matrix $\boldsymbol{U}$.

Given the graph bias term $b(\boldsymbol{h}_{n-1},\boldsymbol{y}_n, \boldsymbol{y}_k)$, the node propagation probability of GBTPP model is given by
\begin{align}\label{eq:P}\notag
&P\left(v_{n+1}=k | \boldsymbol{h}_{n-1}, \boldsymbol{y}_{n}\right) \\
=&
\frac{\exp \left(\boldsymbol{V}_{k, :}^{h} \boldsymbol{h}_{n-1} + b(\boldsymbol{h}_{n-1},\boldsymbol{y}_n, \boldsymbol{y}_k) + b_{k}^{h}\right)}
{\sum_{k=1}^{V} \exp \left(\boldsymbol{V}_{k, :}^{h} \boldsymbol{h}_{n-1} + b(\boldsymbol{h}_{n-1},\boldsymbol{y}_n, \boldsymbol{y}_k) + b_{k}^{h}\right)},
\end{align}
where the direct influence of current node $v_n$ is measured by the bias term $b(\boldsymbol{h}_{n-1},\boldsymbol{y}_n, \boldsymbol{y}_k)$ in Eq.\ref{eq:bias}, and the indirect influence of propagation history is computed using the history embedding vector $\boldsymbol{h}_{n-1}$.

For conventional intensity function, the direct influence of current node $v_n$ is also measured by a separate bias term using node embedding $\boldsymbol{y}_n$ as
\begin{align}\label{eq:lambda}\notag
\lambda^{*}(t) =& \exp \Bigg(
\underbrace{\boldsymbol{v}^{h \top} \cdot \boldsymbol{h}_{n-1}}_{\text { history influnce }} + \underbrace{\boldsymbol{v}^{y \top} \cdot \boldsymbol{y}_{n}}_{\text { direct influence }}\\ & + \underbrace{w^{t}\left(t-t_{j}\right)}_{\text { exponential assumption }} + \underbrace{b^{t}}_{\text { base intensity }}
\Bigg),
\end{align}
where $v^h$, $v^y$ are column vectors, and $w^t$, $b^t$ are scalars.
We list the specific meaning of the terms computed in Eq.\ref{eq:lambda}, and the same term is also used in conventional TPP in Eq.\ref{eq:conv_lambda} except for the \emph{direct influence} term. Specifically,
\begin{itemize}
	\item The \emph{history influence} term $\boldsymbol{v}^{h \top} \cdot \boldsymbol{h}_{n-1}$ represents the accumulative influence from the history nodes and the timing information of
	the past propagation.
	\item The \emph{direct influence} term $\boldsymbol{v}^{y \top} \cdot \boldsymbol{y}_{n}$ represent the influence current node $v_n$.
	\item The \emph{exponential assumption} term $w^{t}\left(t-t_{j}\right)$ assume that the intensity is an exponential function of $t$, where the exponential function acts as a non-linear transformation and guarantees that the intensity is positive.
	\item The last \emph{base intensity} term gives a base intensity level for the propagation process.
\end{itemize}
Based on the conditional intensity function $\lambda^{*}(t)$, we can derive the likelihood that the event propagates from $v_n$ to $v_{n+1}$ at the time $t$ given the history $\boldsymbol{h}_{n-1}$ by the following equation:
{\small
	\begin{align}\label{eq:ll_t}\notag
	f^{*}(t)=&\lambda^{*}(t) \exp \left(-\int_{t_{n-1}}^{t} \lambda^{*}(\tau) d \tau\right) \\\notag
	\notag = &\exp \Bigg\{\boldsymbol{v}^{t^{\top}} \cdot \boldsymbol{h}_{n-1} + \boldsymbol{v}^{y \top} \cdot \boldsymbol{y}_{n} + w^{t}\left(t-t_{n-1}\right)+b^{t} \\\notag
	& -\frac{1}{w} \exp \left(\boldsymbol{v}^{t \top} \cdot \boldsymbol{h}_{n-1} + \boldsymbol{v}^{y \top} \cdot \boldsymbol{y}_{n} + w^{t}\left(t-t_{n-1}\right) + b^{t}\right)  \\
	& +\frac{1}{w} \exp \left(\boldsymbol{v}^{t^{\top}} \cdot \boldsymbol{h}_{n-1}+b^{t}\right)
	\Bigg\}.
	\end{align}
}
\subsubsection{Prediction} For propagation node prediction, given the propagation probability in Eq.\ref{eq:P}, the next propagation node $\hat{v}_{n+1}$ is given by
\begin{equation}\label{eq:node}
\hat{v}_{n+1} = \arg\max_{v_{n+1}\in V} P\left(v_{n+1}| \boldsymbol{h}_{n-1}, \boldsymbol{y}_{n}\right),
\end{equation}
where $V$ is the node set for graph $\mathcal{G}=(V, E)$. 

For propagation time prediction, given the time likelihood in Eq.\ref{eq:int}, the predicted propagation time $\hat{t}_n$ from node $v_n$ to the next node is given by 
\begin{equation}\label{eq:int}
\hat{t}_{n}=\int_{t_{n-1}}^{\infty} t \cdot f^{*}(t) d t.
\end{equation}
Commonly the integration in Eq.\ref{eq:int} does not have analytic solutions. A numerical integration technique\cite{seiler1989numerical} for one-dimensional function is used to compute Eq.\ref{eq:int}.

\subsection{Learning Algorithm}
Given a collection of event propagation sequences $\mathcal{C}=\left\{\mathcal{S}^{i}\right\}$, where $\mathcal{S}^{i} = \left(\left(t_{j}^{i}, v_{j}^{i}\right)_{j=1}^{n_{i}}\right)$, the GBTPP model is learned by maximizing the joint log-likelihood given as
{\small
	\begin{equation}\label{eq:ll}
	\sum_{i} \sum_{j}\left(\log P\left(v_{j+1}^{i} | \boldsymbol{h}_{j-1}, \boldsymbol{y}_{j}\right) +
	\log f\left(t_{j}^{i} | \boldsymbol{h}_{j-1}, \boldsymbol{y}_{j}\right)\right),
	\end{equation}
}
where the node propagation probability $P\left(v_{j+1}^{i} | \boldsymbol{h}_{j-1}, \boldsymbol{y}_{j}\right)$ is computed by Eq.\ref{eq:P} and the propagation time likelihood $f\left(t_{j}^{i} | \boldsymbol{h}_{j-1}, \boldsymbol{y}_{j}\right)$ is computed by Eq.\ref{eq:ll_t}.

To optimize the log-likelihood in Eq.\ref{eq:ll}, we implement Back Propagation Through Time (BPTT) to train GBTPP model. Specifically, supposing the size of BPTT is $b$ and the model in Fig.~\ref{fig:alg} is unrolled by $b$ steps, then for each training iteration, $b$ consecutive samples $\left\{\left(t_{k}^{i}, v_{k}^{i}\right)_{k=j}^{j+b}\right\}$ is offered to apply the feed-forward operation through the network. After we unroll the model for b steps through time, all the parameters are shared across these copies, and will be updated sequentially in the back propagation stage with respect to the loss function.

\section{Experiments}
We evaluate our GBTPP on a synthetic dataset and two real-world datasets, and compare it to both discrete-time and continuous-time sequential models, including Recurrent Marked Temporal Point Process (RMTPP)~\cite{du2016recurrent}. Empirical results show that the GBTPP model achieves better performance on both propagation node prediction and time prediction.

\subsection{Baselines}
For evaluating predictive performance of forecasting propagation node, we compare GBTPP with the following discrete-time models, including:
\begin{itemize}
	\item \textbf{Majority Prediction} For each time when making predictions, we always choose the most popular propagation node by frequency count based on all propagations through current node $v_n$, regardless of propagation history. This is also known as the $1$-order Markov Chain (MC-$1$).
	\item \textbf{Markov Chain} We also compare with Markov models with higher order, including $2$-order and $3$-order denoted as MC-$2$ and MC-$3$ respectively. Instead of considering only $v_n$, previous propagation node $v_{n-1}$ and $v_{n-2}$ is also taken in.
\end{itemize}

For evaluating the performance of predicting propagation time, we compare with several conventional classical point process models, including:
\begin{itemize}
	\item \textbf{Homogeneous Poisson Process (PP)\cite{kingman1993poisson}} In homogeneous Poisson Process, the inter-event times are independent and identically distributed random variables conforming to the exponential distribution. The conditional intensity function $\lambda^{*}(t) = \lambda_{0}$ is a constant over time and independent of the history $\mathcal{H}_t$, producing an estimate of the average inter-event gap.
	\item \textbf{Hawkes Process (HP)\cite{HawkesBiometrika71}} As aforementioned in related work, Hawkes Process is denoted as
	$$\lambda^{*}(t)=\gamma_{0}+\alpha \sum_{t_{j}<t} \gamma\left(t, t_{j}\right),$$
	where $\gamma\left(t, t_{j}\right) \geqslant 0$ is the triggering kernel measuring temporal dependency, $\gamma_{0} \geqslant 0$ is base intensity independent of the history and the summation of kernel terms is history influence. The kernel function can be chosen in advance, e.g.,$\gamma\left(t, t_{j}\right)=\exp \left(-\beta\left(t-t_{j}\right)\right)$
	as we used. The intensity function of HP depends on the history up to time $t$. In general, HP is more expressive than Poisson Process as the events in past intervals can affect the occurrence of the events in later intervals.
	\item \textbf{Self-Correcting Process (SCP)} The Self-Correcting Process is denoted as
	$$\lambda^{*}(t)=\exp \left(\mu t-\sum_{t_{i}<t} \alpha\right),$$
	where $\mu>0, \alpha>0$. Compared with HP, SCP seeks to produce regular temporal patterns. Though the intensity increases steadily, each time a new event appears, the conditional intensity is decreased by multiplying a constant $e^{-\alpha}<1$.
\end{itemize}

We also compare with \textbf{Continuous-Time Markov Chain (CTMC)} model that can jointly predict the node $v_{n+1}$ and time $t_{n}$ for the next propagation step. It learns continuous transition rates between two nodes, and make predictions on the next propagation node with the earliest transition time.

Finally, we compare with state-of-the-art method \textbf{Recurrent Marked Temporal Point Process (RMTPP)}. Similar to the proposed GBTPP model, when dealing with history influence, the temporal dynamic propagation series are embedded into a history vector by recurrent neural network. The major difference between GBTPP and RMTPP lies in that, the structural information of the graph is leveraged in GBTPP through the node vector learned by graph representation, while in RMTPP the event propagation sequence is viewed as general marked event sequences.

Moreover, to further explore the utility of the graph bias term in the GBTPP model and investigate the usage of the node embedding vector by an ablation study, we compare with a \textbf{Node-specific Recurrent Point Process (NRPP)} in which the graph bias term is removed from GBTPP and the node embedding vector only serve as an event profile feature as in the history embedding part in GBTPP. 
Similar approach is firstly used in \cite{yan2016modeling} where the event profile feature is incorporated as a regression prior in conventional multidimensional Hawkes processes. In NRPP and the history embedding part in GBTPP, the embedding vector of the specific node can be seen as a profile feature of the event. It is also a heuristic way to incorporate graph structural information when we first deal with this problem. 

\begin{table*}[tb!]
	\centering
	\resizebox{\textwidth}{!}{
		\begin{tabular}{c|ccccccc|ccccccc}
			\toprule
			$\text{Model}$& MC-1 & MC-2 & MC-3 & CTMC & RMTPP & NRPP & GBTPP & PP & HP & SCP & CTMC & RMTPP & NRPP & GBTPP  \\ \midrule
			$\text{Metrics}$ & \multicolumn{7}{c|}{$\textbf{Accuracy(\%)}$ } & \multicolumn{7}{c}{$\textbf{RMSE}$ } \\\midrule
			\multirow{2}{*}{\scriptsize{Synthetic}}& 17.46 & 25.27 & 33.74 & 32.08 & 46.82 & 46.63 & \textbf{47.26} & 3.457 & 2.164 & 2.845 & 3.420 & 1.852 & 1.905 & \textbf{1.728} \\
			&(2.24) &(2.53) &(1.87) &(2.74) &(1.38) & (1.56) &(1.55) &(0.374) &(0.283) &(0.317) &(0.265) & (0.241) & (0.256) &(0.228) \\\midrule
			\multirow{2}{*}{Higgs}& 10.92 & 14.60 & 16.35 & 17.41 & 22.26 & 22.53 & \textbf{24.59} & 3.267 & 2.518 & 2.343 & 2.355 & 1.741 & 1.622 & \textbf{1.396} \\
			&(2.06) &(1.44) &(1.73) &(2.58) &(1.80) & (1.64) &(1.29) &(0.381) &(0.346) &(0.369) &(0.335) & (0.272) & (0.275) &(0.264) \\\midrule
			\multirow{2}{*}{Meme}& 15.72 & 20.05 & 22.93 & 25.56 & 32.14 & 32.75 & \textbf{35.82} & 2.361 & 1.958 & 1.484 & 1.762 & 1.059 & 0.979 & \textbf{0.825} \\
			&(2.21) &(2.14) &(1.59) &(2.17) &(1.52) & (1.97)  &(1.73) &(0.412) &(0.368) &(0.276) &(0.347) &(0.254) & (0.242) &(0.227)\\\midrule
			
	\end{tabular}}
	\vspace{2pt}
	\caption{
		Mean and standard deviation (in bracket) of metrics for Propagation Node Prediction Accuracy(\%) and Propagation Time Prediction Root Mean Square Error (RMSE), on Synthetic data generated by multidimensional Hawkes process, real-world Higgs dataset of tweet propagation on Twitter, and real-world MemeTracker dataset of information propagation between websites. On both synthetic and real-world datasets, GBTPP model outperforms the baselines considering Accuracy and RMSE. 
		}\label{tab:res}
\end{table*}

\subsection{Dataset}
To verify the potential of the proposed model, we evaluate its performance on two real-world datasets, including a Higgs Twitter dataset\cite{de2013anatomy} to predict retweeting between users and a MemeTracker dataset\cite{Leskovec:2009:MDN:1557019.1557077} to predict meme propagation between websites.

\textbf{Synthetic} To simulate event propagation processes in graph e.g. user activities in social networks, we use a multi-dimensional Hawkes process to generate a synthetic dataset. Hawkes process is widely used to model the generative process of user behavior in social networks, like \cite{li2014learning, li2013dyadic, ZhouAISTATS13}.
To generate event propagation sequences in a graph with $U$ nodes, we set $U$ Hawkes processes that are coupled with each other: each of the Hawkes processes corresponds to an individual node and the influence between nodes are explicitly modeled
\footnote{Here the multi-dimensional Hawkes process belongs to generative models generate sequences statistically similar to social activities, which is different from predictive models like the proposed GBTPP and state-of-the-art RMTPP. One can see \cite{xu2017patient}  for a detailed specification for generative models and predictive/discriminative models.}.
Specifically, the multi-dimensional Hawkes process is defined by a $U$-dimensional point
process $N_{t}^{u}, u=1, \ldots, U$, with the conditional intensity
for the $u$-th dimension defined as:
$$
\lambda_{u}(t)=\mu_{u}+\sum_{i : t_{i}<t} a_{u u_{i}} g\left(t-t_{i}\right),
$$
where $\mu_{u} \geq 0$ is the base intensity for the $u$-th Hawkes process, $a_{u u^{\prime}} \geq 0$ captures the mutually-exciting influence between the $u$-th and $u'$-th node.
Larger value of $a_{u u^{\prime}}$ indicates that events are more likely to propagate from the $u'$-th node to the $u$-th node in the future. We collect the parameters into matrix-vector forms with $\boldsymbol{\mu}=\left(\mu_{u}\right)$ for the base intensity, and $\mathbf{A}=\left(a_{u u^{\prime}}\right)$ for the mutually exciting coefficients called infectivity matrix.

In this experiment, we set $U=100$ and generate propagation sequences with randomly initialized parameter $\mathbf{A}$ and $\boldsymbol{\mu}$. Similar to \cite{ZhouAISTATS13}, the base intensity parameters $\boldsymbol{\mu}$ are sampled from uniform distribution on $[0, 0.001]$, and the infectivity matrix $\mathbf{A}$ is generated by $\mathbf{A}=\mathbf{U} \mathbf{V}^{T}$, where $\mathbf{U}$ and $\mathbf{V}$ are both $100\times 9 $ matrices with entries $[10(i-1)+1 : 10(i+1), i ], i=1, \ldots, 9$ sampled randomly from $[0,0.1]$ and all other entries are set zero.
Then we scale $\mathbf{A}$ so that the spectral radius of $\mathbf{A}$ is 0.8 to ensure the point process is well-defined with finite intensity. In the end, we sample 50,000 sequences from the multi-dimensional Hawkes process specified by $\mathbf{A}$ and $\boldsymbol{\mu}$ for the training and testing of baselines and proposed GBTPP model by 10-fold cross validation.

\textbf{Higgs} The Higgs dataset is a public dataset built by monitoring the spreading processes on Twitter before, during and after the announcement of the discovery of a new particle with the features of the elusive Higgs boson on 4th July 2012. Messages between 1st and 7th July 2012 about this discovery posted in Twitter are considered.
There are four directional networks available in the dataset based on user activities, including a retweet network (retweeting between users), a reply network (replying to existing tweets), a mention network (mentioning other users) and a social network (friends/followers social relationships among user involved in the above activities).
In the experiment, we study the tweet propagation process using the largest strongly connected component in the directed and weighted retweet network with 984 nodes (users) and 3,850 edges. 
Firstly a graph embedding $\{\boldsymbol{y}_k\}_{k=1}^V$ is learned by graph representation where $\boldsymbol{y}_k$ is the embedded node vector for node $v_k$, then the GBTPP model is trained on the retweet activities.

\textbf{Meme} The MemeTracker dataset is also a public dataset which is widely in TPP works \cite{XiaoNIPS17,ZhouAISTATS13,mei2017neural}. The dataset contains the information flows captured by hyper-links between different sites with timestamps. It tracks meme diffusion over public media, containing more than 172 million news articles or blog posts. The memes are sentences, such as ideas, proverbs, and the time is recorded when it spreads to certain websites.
In the experiment, we extract the top 500 popular sites and 62,593 meme propagation cascades between them. Firstly, the adjacent matrix is estimated by $A_{i,j}=\frac{N_{ij}}{N_{max}}$, where $N_{ij}$ is the number of observed meme propagations from website $v_i$ to $v_j$ in the propagation cascades and $N_{max}$ is the global maximum number of meme propagations between websites. Given adjacent matrix $A_{i,j}$, the graph embedding $\{\boldsymbol{y}_k\}_{k=1}^V$ is learned where $\boldsymbol{y}_k$ is the embedded node vector for website $v_k$, then the GBTPP model is trained on meme propagation cascades.

Our experiments are conducted under Ubuntu 64bit 16.04LTS, with i5-8600K 3.60GHz$\times$6 CPU, 16G RAM and NVIDIA GeForce GTX 1070Ti GPU. All the experimental results are given by 10-fold cross validation. It is worth mentioning that, it is required to maintain the integrity of propagation process data sample when we train and test our model. So instead of deleting a chunk of the propagation sequence and train on the rest, we consider a sequence as one complete and independent sample when dividing the whole dataset into $10$ subsets to perform 10-fold cross validation.

To elaborate the 10-fold cross validation implementation:
Given a collection $\mathcal{C}$ of observed event propagation sequences $\mathcal{C}=\left\{\mathcal{S}^{1}, \mathcal{S}^{2}, \ldots\right\}$, where $\mathcal{S}^{i}$ is the $i$-th propagation sequence $\mathcal{S}^{i}=\left(\left(t_{1}^{i}, v_{1}^{i}\right),\left(t_{2}^{i}, v_{2}^{i}\right), \ldots\right)$ recording the event propagation process in graph $\mathcal{G}=(V, E)$, the collection $\mathcal{C}$ is randomly partitioned into $10$ equal sized subsets $\mathcal{C}_1$ -- $\mathcal{C}_{10}$. Then the 10-fold cross validation can be implemented: Of the $10$ subsets $\mathcal{C}_1$ -- $\mathcal{C}_{10}$, one single subset is retained as the validation data for testing the model, and the remaining $9$ subsets are used as training data.

Specifically, we then prepare training or testing samples from the training or validation dataset for our model, including input data and output label. For a sequence e.g. of length $N$, $\mathcal{S}=\left(\left(t_{1}, v_{1}\right),\left(t_{2}, v_{2}\right), \ldots, \left(t_{N}, v_{N}\right) \right)$ in the training or validation data, we acquire $N-1$ training or testing samples as we observe $N-1$ propagations between nodes, e.g, the propagation from form node $v_n$ to $v_{n+1}$, the model \emph{input} include propagation history $\mathcal{H}_{n-1}=\big((t_{1}, v_{1}),(t_{2}, v_{2}), \dots, (t_{n-1}, v_{n-1})\big)$ and current node $v_n$, and the \emph{output} label is node $v_{n+1}$ and propagation time $t_n$.

\begin{figure}[h]
	\centering
	\subfigure{\includegraphics[width=0.48\textwidth]{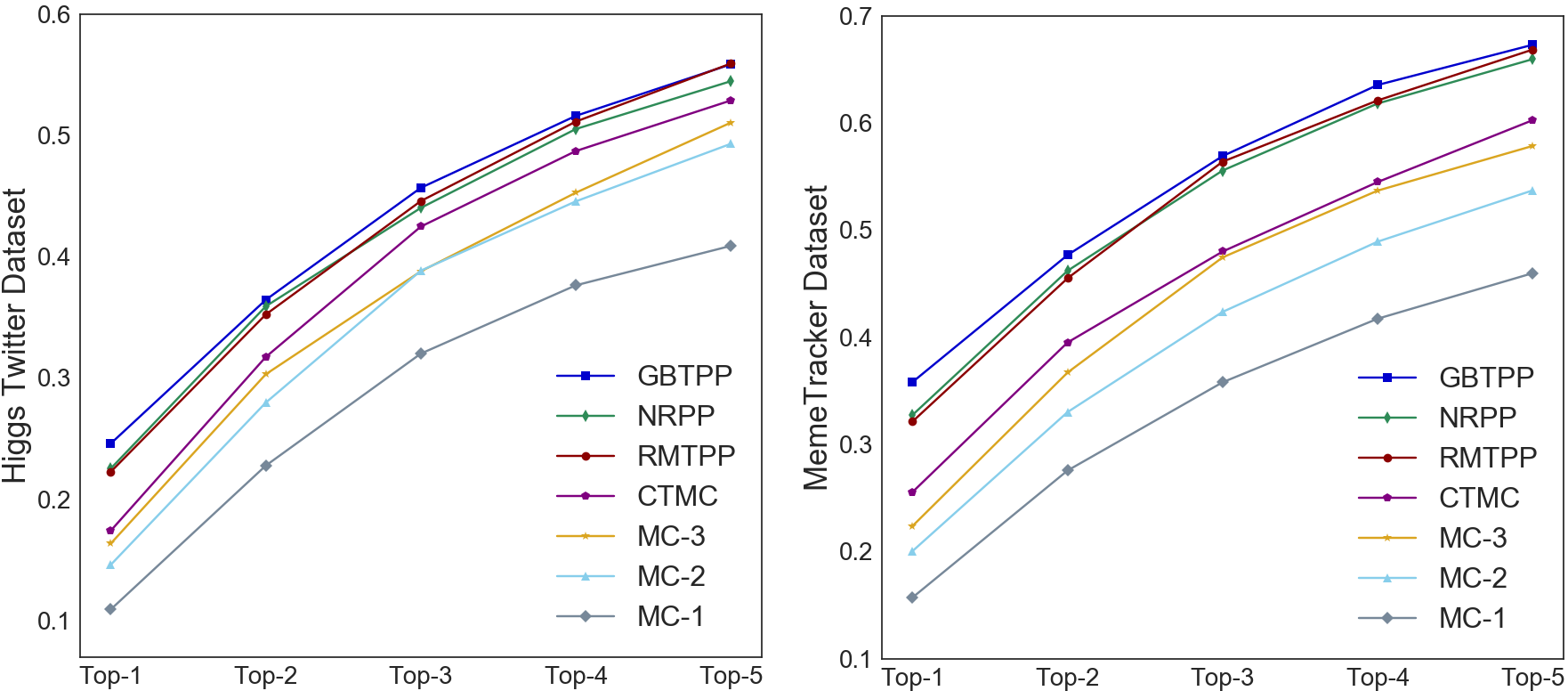}}
	\vspace{-5pt}	
	\caption{Top-5 node prediction accuracy on Higgs dataset and MemeTracker dataset.}
	\label{fig:topk}
	\vspace{-5pt}
\end{figure}

\subsection{Experimental Results}
We use prediction accuracy (\# correct predictions divide total predictions) to evaluate propagation node prediction, and Root Mean Square Error (RMSE) to evaluate propagation time prediction. The empirical results with standard deviation are presented in Table.\ref{tab:res}.

Moreover, we further compute the top-$K$ precision curve for propagation node prediction on Higgs Twitter dataset and MemeTracker dataset in Fig.~\ref{fig:topk}. Top-$K$ precision curve is widely used for recommender systems. In fact, our model can act as a recommender system recommending next propagation node $\hat{v}_{n+1}$ in the period of propagation time $\hat{t}_n$, e.g., recommending interested tweets for user $v_{n+1}$ at the time around $t_n$, or recommending popular news and memes to the editors of website $v_{n+1}$ around time $t_{n}$.

Specifically, we have the following findings and discussions:

i) \textbf{Recurrent Model vs Parametric Model}
As shown in Table.\ref{tab:res}, GBTPP and RMTPP outperforms conventional parametric methods like Markov Chain and point process models like Poisson Process, HP and SCP, as well as joint prediction model CTMC on both propagation node prediction and time prediction. The main advantage lies in that, conventional models make strong assumptions on the distribution form and generative process of the data, while GBTPP and RMTPP use recurrent neural networks to automatically learn the influences from propagation history.

ii) \textbf{RMTPP vs GBTPP}
Compared with state-of-the-art RMTPP model, the GBTPP model achieves better performance than RMTPP, especially on real-world datasets.
Though we use a multi-dimensional Hawkes process to simulate event propagation sequences, the actual graph structure can not be simulated like real-world dataset. Correspondingly, it explains to the results that the GBTPP model achieves comparably better performance than RMTPP on real-world datasets than the synthetic one.

Two major innovations contribute to the promotion of GBTPP model compared with RMTPP model: i) The structural information of the graph is used in the form of node embedding as side information. ii) The direct influence between connected node is separately measured as an extra bias term from indirect influence of propagation history. It verifies our hypothesis that as a special case of event sequence modeling, event propagation modeling in graph requires more suitable model to deal with the structural information and reflect the fact that event propagation is more likely to happen between the connected nodes in graph.

iii) \textbf{NRPP vs GBTPP} As shown in Table. \ref{tab:res} and Figure. \ref{fig:topk}, while NRPP model achieve comparable results against RMTPP as it incorporate node embedding vectors as node-specific profile feature, the GBTPP model still have distinct superiority over NRPP on both numerical results and top-$K$ precision curve. This empirical results suggest that, though the better performance of GBTPP model benefits from introducing graph structural information, the utility method e.g. modeling direct and indirect influence using graph bias term in GBTPP is more important. The improvement is quite limited only by introducing graph embedding vector and higher network complexity.

\section{Conclusion}

Temporal point processes are widely used for modeling event sequences, while event propagation sequence modeling is rarely considered as a special case, where the structural information and direct connections between nodes are not utilized.
In this paper, we study the problem of event propagation modeling by Graph Biased Temporal Point Process. Compared with state-of-the-art method, we have two innovations: i) The direct influence between connected nodes is separately measured as an extra bias term from indirect influence of the propagation history, through pre-learned graph representation. ii) When modeling the indirect influence of the propagation history, the structural information of the graph is used in the form of node embedding as side information. We evaluate GBTPP model on Higgs Twitter dataset predicting retweeting in social network and MemeTracker dataset predicting meme propagation between websites. Experimental results collaborate the effectiveness of our approach compared to conventional methods and state-of-the-art method.

As a future work, the GBTPP can be extended to modeling event propagation in dynamic networks by introducing graph embedding method e.g.\cite{singer2019node} over dynamic temporal graphs. In addition, the semantic information of the event e.g. text information of news and tweets if it is available, can be incorporated into the model to improve the performance. 

\section{Acknowledgement}
The work is partly supported by National Key Research and Development Program of China (2016YFB1001003), STCSM (18DZ1112300), and the Interdisciplinary Program of Shanghai Jiao Tong University (project number ZH2018QNB12).
\bibliographystyle{abbrv}
\bibliography{sample-bibliography}

\vspace{-35pt}
\begin{IEEEbiography}[{\includegraphics[width=1in,height=1.25in,clip,keepaspectratio]{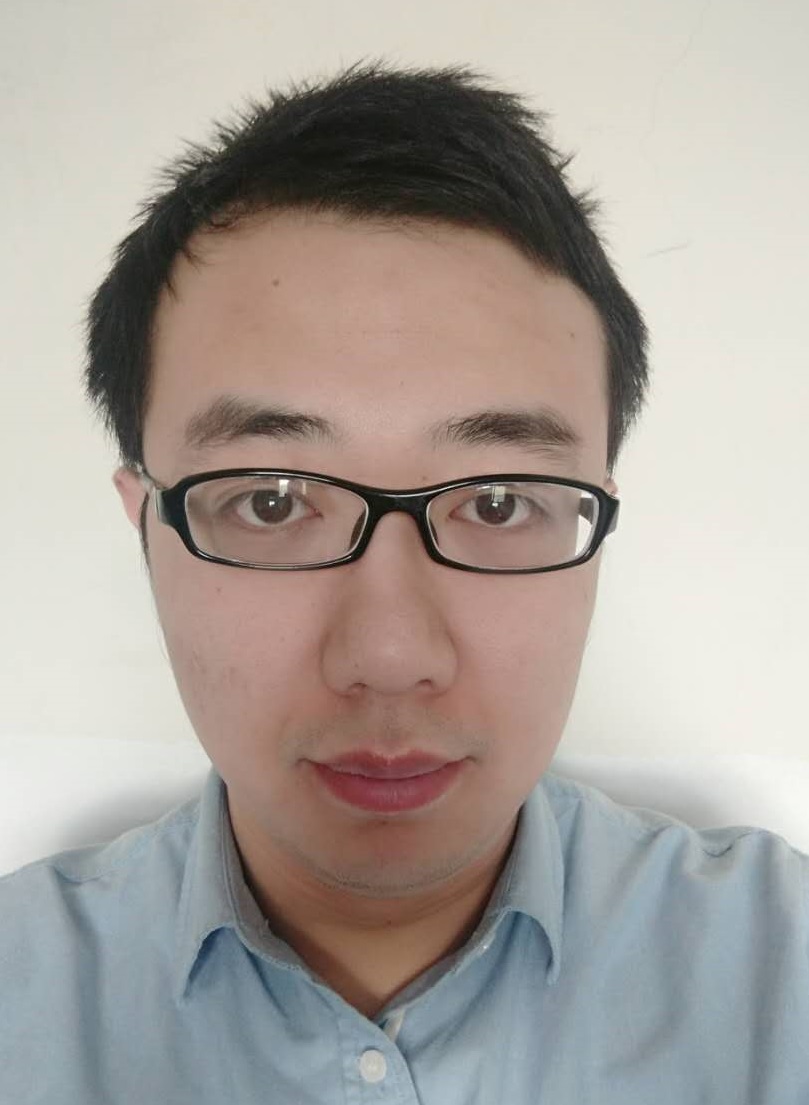}}]{Weichang Wu}
	received the B.S. degree in electronic engineering from Huazhong University of Science and Technology, Wu Han, China, in 2013. He is currently pursuing the Ph.D. degree with the Department of Electronic Engineering, Shanghai Jiao Tong University, Shanghai, China. His current research interests include machine learning and data mining, especially sequential data analysis and temporal point processes.
\end{IEEEbiography}
\vspace{-35pt}
\begin{IEEEbiography}[{\includegraphics[width=1in,height=1.25in,clip,keepaspectratio]{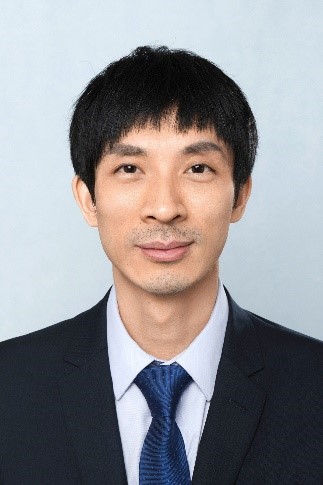}}]{Huanxi Liu}
	was born in Hunan, China, in 1982. He received his Master and Doctor degrees in Pattern Recognition and Intelligent System from Shanghai Jiao Tong University, China, in 2007 and 2010, respectively. He is currently an Senior Engineer at the Institute of Image Processing and Pattern Recognition, Shanghai Jiao Tong University, China. His research interests include pattern recognition, computer vision and machine learning.
\end{IEEEbiography}
\vspace{-35pt}
\begin{IEEEbiography}[{\includegraphics[width=1in,height=1.25in,clip,keepaspectratio]{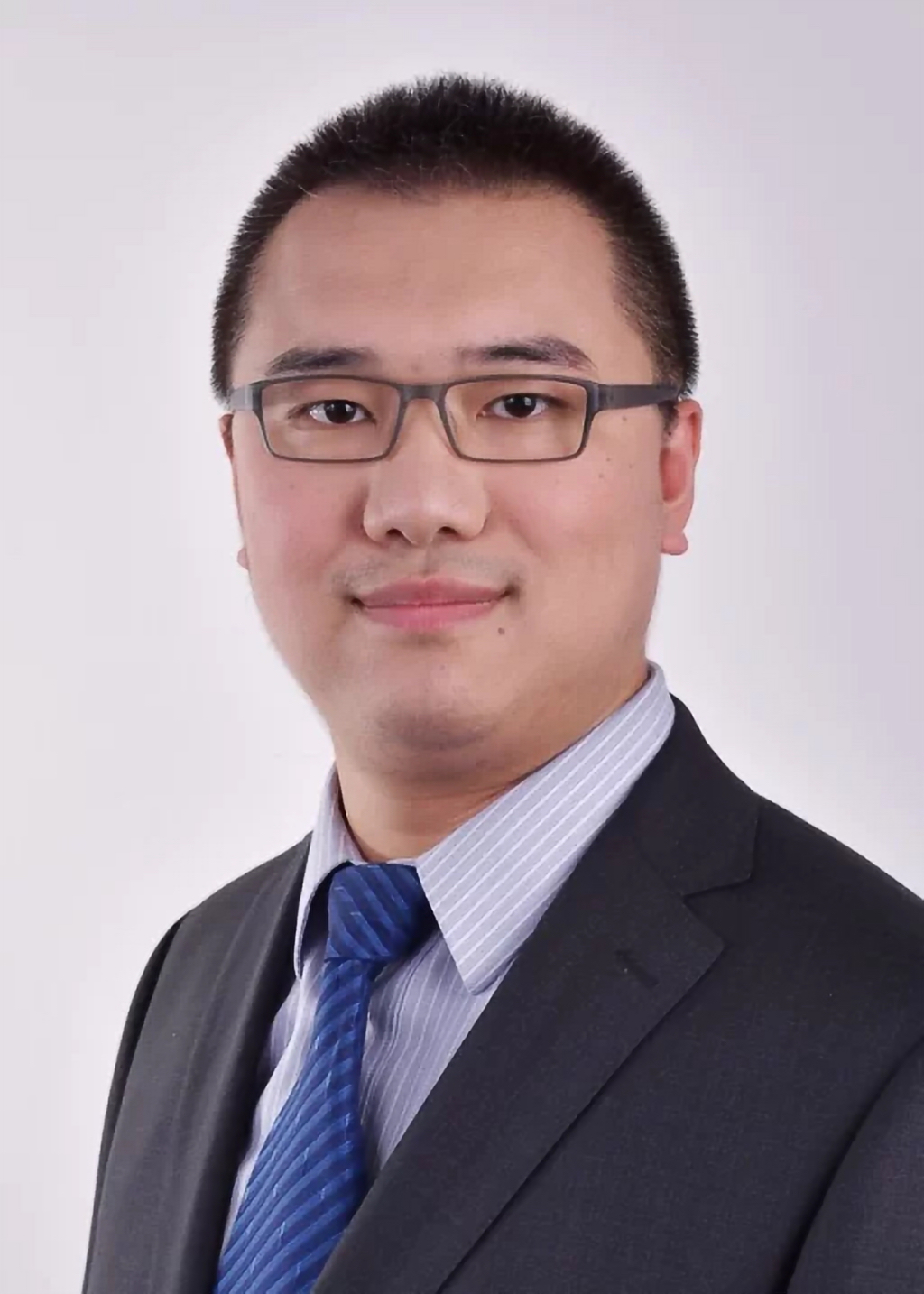}}]{Xiaohu Zhang}
	graduated from the Software Engineering, Northwest University in 2012 with a master's degree. Now he is Senior Director, the head of Innovation Center in China Telecom BestPay Co., Ltd. In 2016, he is responsible for leading the implementation of Orange Cloud Cutting, China Telecom BestPay. He won the "Special Contribution Award" which is used to commend him for his special contributions to the company in intelligent operation and maintenance, block chains, system architecture, etc in 2018.
\end{IEEEbiography}
\vspace{-35pt}
\begin{IEEEbiography}[{\includegraphics[width=1in,height=1.25in,clip,keepaspectratio]{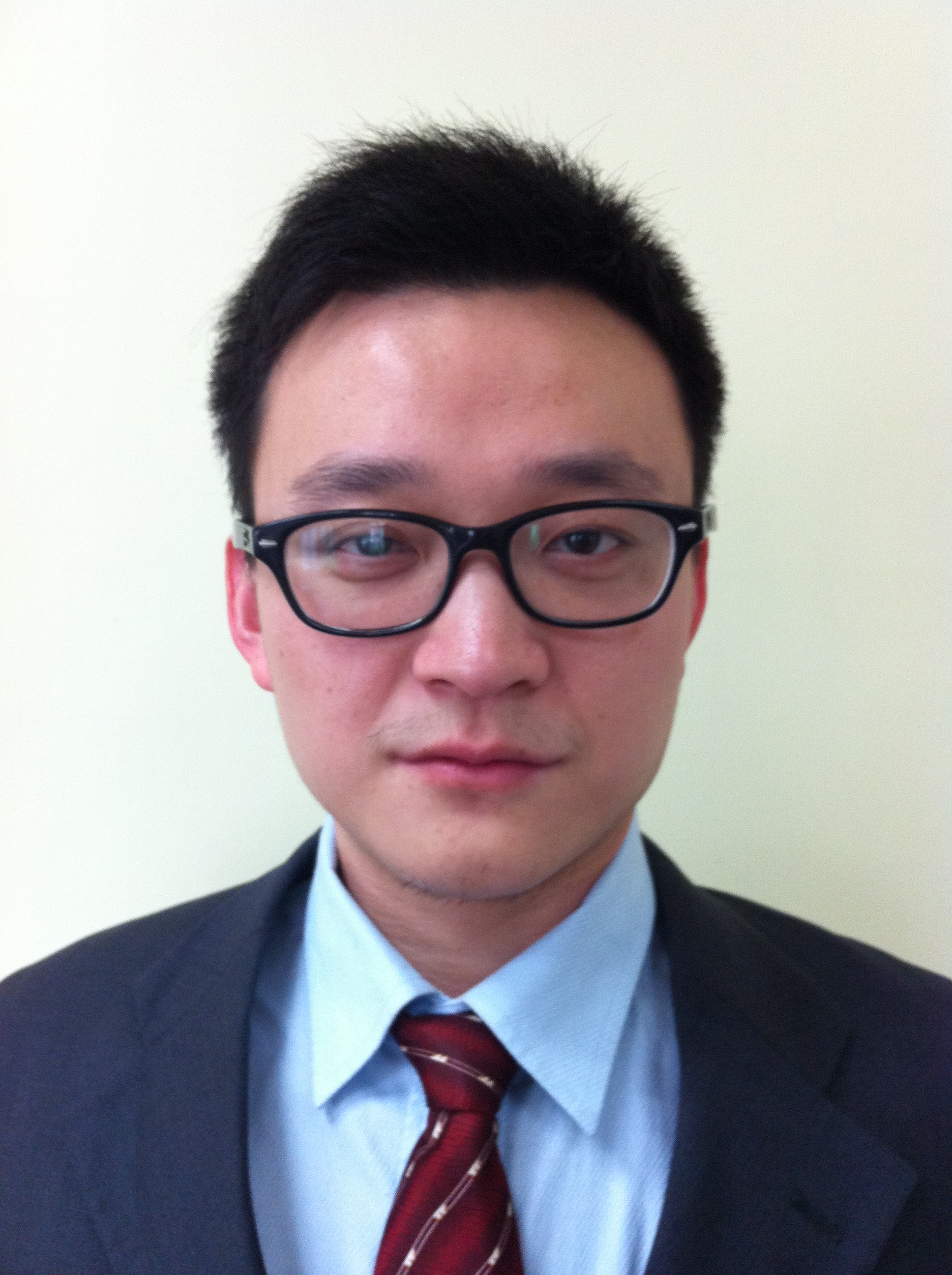}}]{Yu Liu}
	graduated from the Communication and Information System, Donghua University in 2009 with a master's degree. Now he is a senior engineer in China Telecom BestPay Co., Ltd. He has published in DBAplus "Building an automated database monitoring system with Zabbix and Ansible", "Redis retrieval performance is insufficient, and reshaping rsbeat to solve historical slow log tracking".
\end{IEEEbiography}
\vspace{-35pt}
\begin{IEEEbiography}[{\includegraphics[width=1in,height=1.25in,clip,keepaspectratio]{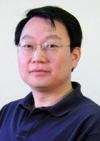}}]{Hongyuan Zha} is a Professor at the School of Computational Science and Engineering, College of Computing, Georgia Institute of Technology and adjunct Professor with Shanghai Jiao Tong University and East China Normal University. He earned his PhD degree in scientific computing from Stanford University in 1993. Since then he has been working on information retrieval, machine learning applications and numerical methods. He is the recipient of the Leslie Fox Prize (1991, second prize) of the Institute of Mathematics and its Applications, the Outstanding Paper Awards of the 26th International Conference on Advances in Neural Information Processing Systems (NIPS 2013) and the Best Student Paper Award (advisor) of the 34th ACM SIGIR International Conference on Information Retrieval (SIGIR 2011). He was an Associate Editor of IEEE Transactions on Knowledge and Data Engineering.
\end{IEEEbiography}

\end{document}